\documentclass[prd,twocolumn,nofootinbib]{revtex4}
\usepackage{graphicx, epsfig}
\usepackage{color}
\usepackage{mathrsfs}
\usepackage {amssymb}
\usepackage {amsmath}

\input epsf
\usepackage{color,graphics,latexsym,amsfonts}

\begin{document}
\title{Relativistic stars in f(R) gravity}
\author{E. Babichev, D. Langlois}
\affiliation{APC, UMR 7164 (CNRS-Universit\'e Paris 7), 10 rue Alice Domon et L\'eonie Duquet,75205 Paris Cedex 13, France}
\begin{abstract}
We study the strong gravity regime in viable models of so-called $f(R)$  gravity that account for the observed cosmic acceleration.  In contrast with recent works suggesting that  very relativistic stars  might not exist in  these models, we find numerical solutions corresponding to  static star configurations with a strong gravitational field. The choice of the equation of state for the star is crucial for the existence of solutions. 
Indeed, if the pressure exceeds one third of the energy density in a large part of the star,  static configurations do not exist.
In our analysis, we use a polytropic equation of state, which is not plagued with this problem and, moreover, provides a better approximation for a realistic neutron star.

\end{abstract}

\date{\today} 
\maketitle

\def\beq{\begin{equation}}
\def\eeq{\end{equation}}
\def\be{\begin{equation}}
\def\ee{\end{equation}}
\newcommand{\bea}{\begin{eqnarray}}
\newcommand{\eea}{\end{eqnarray}}
\def\bi{\begin{itemize}}
\def\ei{\end{itemize}}
\def\ba{\begin{array}}
\def\ea{\end{array}}
\def\bfig{\begin{figure}}
\def\efig{\end{figure}}

\def\half{\mbox{\scriptsize{${\frac{1}{2}}$}}}
\def\halff{\mbox{\scriptsize{${\frac{5}{2}}$}}}
\def\quarter{\mbox{\scriptsize{${\frac{1}{4}}$}}}
\def\eighth{\mbox{\scriptsize{${\frac{1}{8}}$}}}

\def\mP{{M_P}}  % reduced Planck mass=1/\sqrt{8\pi G}
\def\gs{g_{\rm{s}}}
\def\ls{\ell_{\rm{s}}}
\def\lP{\ell_{\rm{P}}}
\def\ld{\ell_{{d}}}
\def\n{{\bf \hat{n}}}
\def\v{{\bf v}}

\newcommand\bmp[1]{\begin{minipage}[c]{#1\textwidth}}
\newcommand\emp{\end{minipage}}

\def\a{{\alpha}}
\def\d{{\delta}}
\def\e{{\epsilon}}
\def\tg{{\tilde g}}
\def\tR{{\tilde R}}
\def\tH{{\tilde H}}
\def\tT{{\tilde T}}
\def\tn{{\tilde n}}
\def\trho{{\tilde\rho}}
\def\tP{{\tilde P}}
\def\hphi{{\hat\phi}}
\def\hrho{{\hat \rho}}
\def\hP{{\hat P}}
\def\hV{{\hat V}}
\def\hm{{\hat m}}
\def\M{{\cal M}}
\def\hM{{\hat{\cal M}}}
\def\F{{\cal F}}
\def\V{{\cal V}}
\def\dphi{{\delta\phi}}
\def\phim{{\phi_{\rm min}}}

\def\R0{{R_0}}
\def\x1{x_\infty}

\par\bigskip

One of the most challenging tasks for cosmology and fundamental physics today is to try to understand the apparent acceleration of the Universe. Among the many models which have been proposed, a class which has attracted a lot of attention is the so-called $f(R)$ gravity theories where the standard Einstein-Hilbert gravitational Lagrangian, proportional to the scalar curvature $R$, is replaced by a function of $R$ while the matter part of the Lagrangian is left unchanged (see e.g. \cite{review} for a recent review). 
After many detours, it has been realized that viable $f(R)$ theories
must satisfy stringent conditions in order to avoid instabilities and to
be compatible with the present laboratory and astrophysical constraints. A
few models \cite{Hu:2007nk,Appleby:2007vb,Starobinsky:2007hu} have been
carefully constructed to meet these requirements, using in particular  the
chameleon mechanism \cite{KW} to satisfy the solar system constraints (the
binary pulsar constraints are, so far, weaker for these models
\cite{EspositoFarese:2009ta}).

However, recent works \cite{Frolov,KM1} studying the strong gravity limit of these models have questioned their viability by suggesting that, inside neutron stars, where the effects of general relativity are strong, one would easily reach the singularity where $R$ becomes infinite. In particular, in \cite{KM1}, this impression seemed  reinforced by the impossibility to construct numerically relativistic stars beyond some critical value of the gravitational field. 

In this Letter, we reexamine this question and, unlike these previous works, find that highly relativistic stars can be obtained numerically. We also give qualitative arguments to understand the existence of these solutions. 
%Since the $f(R)$ models are equivalent to a specific subclass of scalar-tensor theories, we have also studied highly relativistic stars in chameleon models \cite{KW} and obtained qualitatively similar results. Here we present our main results concerning the $f(R)$ models, while  further details and our results on chameleon models will be discussed elsewhere \cite{BL2}. 
An important conclusion of our work is that the equation of state must satisfy $\rho-3P>0$ in most of the star. Otherwise, tachyonic instabilities associated with a negative effective squared mass develop and prevent the existence of a static star configuration. This problem affects in particular the highly relativistic constant energy density stars. This is why we have used a polytropic equation of state, which is has the additional advantage to be a better approximation to a realistic neutron star.

%\section{Model and equations}
Our starting point is the action
\beq
\label{jordan}
S=\frac{M_P^2}{2}\int d^4x \sqrt{-\tg}\,f(\tR) + S_m[\Phi_m;\tg_{\mu\nu}], 
\eeq
where $S_m$ is the action for the matter, which is minimally coupled to the so-called Jordan metric $\tg_{\mu\nu}$;  $\tR$ is the scalar curvature associated with $\tg_{\mu\nu}$ and we have defined $M_P^{-2}\equiv 8\pi G$. 
In contrast with \cite{Frolov} and \cite{KM1}, we reexpress the model as a scalar-tensor theory in the so-called Einstein frame. 
The two formulations are of course equivalent (at least on the classical level), but the Einstein frame is useful to see that the behaviour of $f(R)$ theories is quite analogous to chameleon-like scalar tensor theories. 

By introducing the scalar field $\phi=\sqrt{\frac{3}{2}} M_P \, \ln f_{,\tR}$ and the metric $g_{\mu\nu}=\Omega^{-2}\, \tg_{\mu\nu}$, with 
$\Omega^{-2}=f_{,\tR}=\exp[\sqrt{\frac23}\, \phi/M_P]\equiv \exp[-2Q\phi/M_P]$, 
the action (\ref{jordan}) becomes 
\begin{eqnarray}
\label{einstein}
S&=& \int d^4x \sqrt{-g}\left[\frac{M_P^2}{2}R
-\frac{1}{2}\left({\nabla}\phi\right)^2-V(\phi)\right] \nonumber\\
&&+{S}_m[\Phi_m;\Omega^2 g_{\mu\nu}],
\end{eqnarray}
with the potential
\beq
V=M_P^2\, \frac{\tR f_{,\tR}-f}{2f_{,\tR}^2},\nonumber
\eeq
which can be expressed in terms of $\phi$ by inverting the definition of  $\phi$ as a function of  $\tR$.

The first and second derivatives of the potential $V$ in terms of 
$\phi$ are
\be
\label{dV}
\begin{aligned}
\frac{dV}{d\phi} &=
\sqrt{\frac{2}{3}}M_P\,  \frac{2f-\tR f_{,\tR}}{2f_{,\tR}^2}\,, \\
\frac{d^2V}{d\phi^2} &=\frac{1}{3f_{,\tR\tR}} 
\left[ 1+ \frac{\tR f_{,\tR\tR}}{f_{,\tR}}-\frac{4ff_{,\tR\tR}}{f_{,\tR}^2} 
\right]\,.
\end{aligned}
\ee

\begin{figure}[ht]
\centering
\includegraphics[width=0.48\textwidth, clip=true]{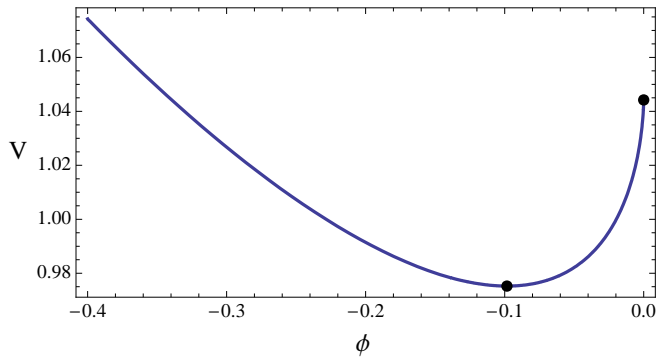}
\caption{Potential $V$ (in units of $M_P^2 \R0$) as a function of $\phi$ (in Planck units) for $n=1$ and $\x1=3.6$.
The lower black dot corresponds to the de-Sitter attractor while the upper-right dot shows the curvature singularity.}
\label{V}
\end{figure}

The function $f(R)$ is severely constrained by observations. Here we will consider the model suggested by Starobinsky \cite{Starobinsky:2007hu},
\beq
\label{staro}
f(\tR)=\R0\left[x-\lambda 
\left(1-\left( 1+x^2\right)^{-n} \right)\right], \quad x\equiv \frac{\tR}{\R0}\, .
\eeq
For this model the potential $V(\phi)$ is shown in Fig.~\ref{V}.
Substituting this expression into (\ref{dV}) yields the minimum of $V$  corresponding to the asymptotic de Sitter solution $\tR_\infty\equiv \x1 \R0$. It is convenient to express the parameter $\lambda$ in terms of $\x1$:
\beq
\lambda=\frac{\x1(1+\x1^2)^{n+1}}
{2\left[ (1+\x1^2)^{n+1}-1-(n+1)\x1^2 \right]}\,.\nonumber
\eeq

In the strong gravity limit, corresponding to $x\gg 1$, we have
\beq
f(\tR)\approx  \R0\left(x-\lambda +\frac{\lambda}{x^{2n}}\right), \quad  \frac{\phi}{\mP}\approx -\frac{\sqrt{6}\lambda n}{ x^{2n+1}} ,
\label{approx}
\eeq
The curvature singularity thus corresponds to a finite value $\phi=0$ for the scalar field, as well as a finite value for the potential 
$V=(\lambda/2)M_P^2 \R0$. However, the derivative $dV/d\phi$ becomes infinite in the limit $\phi\rightarrow 0^-$.

%figure
For matter we consider a perfect fluid characterized by the energy density $\trho$ and pressure $\tP$, defined in the Jordan frame (i.e. with respect to the metric $\tg_{\mu\nu}$). In the Einstein frame, the corresponding energy density and pressure are 
respectively $\rho=\Omega^4\trho$ and $P=\Omega^4\tP$. Inside the star, $\phi$ will be extremely close to zero and therefore, 
the quantities defined in the two frames will be numerically indistinguishable. 

We consider a static and spherically symmetric metric (in the Einstein frame)
\beq
ds^2=-e^\nu dt^2+e^\lambda dr^2+r^2 \left(d\theta^2+\sin^2\!\theta\, d\varphi^2 \right).\nonumber
\eeq
Introducing the radial function $m(r)$ so that 
$
 e^{-\lambda}=1-2m /r,\nonumber
$
the time and radial components of Einstein's equations yield  
\beq
\label{tt}
m'=\frac{r^2}{2M_P^2}\left[\Omega^4 \trho+\frac12 e^{-\lambda}{\phi'}^2+V(\phi)\right],
\eeq
\beq
\label{rr}
\nu'=e^\lambda\left[\frac{2m}{r^2}+\frac{r}{M_P^2}\left(\frac12 e^{-\lambda}{\phi'}^2-V(\phi)\right)+
\frac{r\Omega^4 \tP}{M_P^2}\right].
\eeq
where a prime denotes a derivative with respect to the radial coordinate, and 
one recognizes the total energy density in the brackets on the right hand side of the first equation. 
Instead of the angular component of Einstein's equations, one can use  the conservation of the energy-momentum conservation,
$\tilde\nabla_\mu \tT^\mu_{\ \nu}=0$, which reads
\beq
\label{Bianchi}
\tP'=-\frac12\left(\trho+\tP\right)\left(\nu'+2\frac{\Omega_\phi}{\Omega}\phi'\right),
\eeq
where $\Omega_\phi\equiv d\Omega/d\phi$.
The Klein-Gordon equation for the scalar field reads,
\beq
\label{KG}
\phi''+\left(\frac{2}{r}+\frac12(\nu'-\lambda')\right)\phi'=e^\lambda\left[\frac{dV}{d\phi}+\Omega^3 \Omega_\phi(\trho-3\tP)\right].
\eeq

In order to close the system of equations (\ref{tt}), (\ref{rr}), (\ref{Bianchi}) and (\ref{KG}), one should finally specify the equation of state for 
the matter. In \cite{KM1}, relativistic stars with constant energy density were considered. The advantage is that  the pressure can  
be easily determined analytically in general relativity (i.e. ignoring the backreaction of the scalar field) and is given by
 \beq
  P(r)=\rho_0 \frac{\left(1-{\frac{2GM}{r_*}}\right)^{1/2}-\left(1-{\frac{2GMr^2}{r_*^3}}\right)^{1/2}}{\left(1-{\frac{2GMr^2}{r_*^3}}\right)^{1/2}-3 \left(1-\frac{2GM}{r_*}\right)^{1/2} },\nonumber
  \eeq
where $M$ is the mass of the star and $r_*$ its radius.
However, this implies that, in the innermost regions of the star, the quantity 
$\rho-3 P$, i.e. the opposite of the trace of the energy-momentum tensor, 
 becomes negative as soon as
\beq
-\Phi_*\equiv \frac{GM}{r_*}>\frac{5}{18}.\nonumber
\eeq
In stars with a large region where  $\trho-3\tP<0$, we were unable to find numerically solutions for the scalar field. This is not related to the presence of the curvature singularity, since it  occurs in chameleon type models as well \cite{BL2}. 
The reason is the presence of instabilities, because  the matter
contribution to the effective squared mass  is proportional to
$\trho-3\tP$ and thus becomes negative. Note that a similar effect has
been observed
in \cite{Harada:1997mr} for another class of scalar-tensor theories,
where $\ln\Omega$ depends quadratically on the scalar field .
\begin{figure}[ht]
\centering
\includegraphics[width=0.48\textwidth, clip=true]{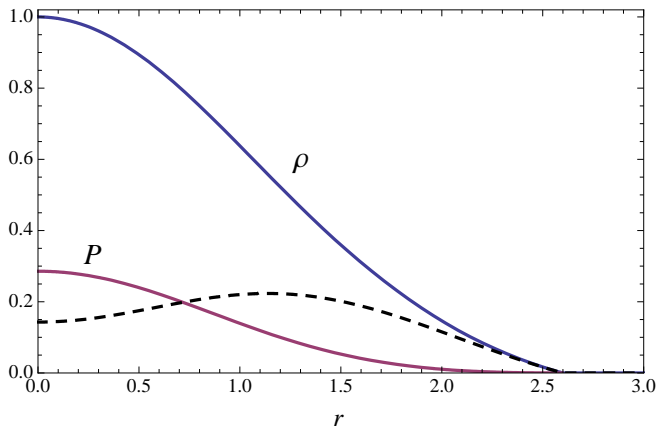}
\caption{Energy density $\trho$ (upper solid line), pressure $\tP$ (lower solid line) and the combination $\trho-3\tP$ (dashed line), 
in units of the central density $\rho_c$,
as functions of the radial coordinate $r$ (in units of $M_P\tilde{\rho}_c^{-1/2}$).}
\label{star}
\end{figure}

The equation of state  deep inside  a neutron star is not known but most realistic equations of state satisfy the condition $\trho-3\tP>0$.  
This is why we have chosen for the present analysis a polytropic  equation of state, given by
\beq
\label{eos}
\trho(\tn)=m_B\left( \tn+K\frac{\tn^2}{n_0}\right) , \quad \tP(\tn)=K m_B\frac{\tn^2}{n_0},
\eeq
with $m_B=1.66\times 10^{-27}$ kg, $n_0=0.1\, {\rm fm}^{-1}$ and $K=0.1$.  
In Fig.~2, the radial profile of the energy density and pressure, together with $\trho-3\tP$, is plotted for a star with the central particle number density $\tn_c = 0.4 \, {\rm fm}^{-3}$. This corresponds to $|\Phi_*|\simeq 0.25$, which is more than twice higher than the value reached in \cite{KM1}.

%\section{Numerical integration and discussion}

We have integrated numerically the system of equations (\ref{tt}-\ref{KG}), together with (\ref{eos}), to obtain the profile of the scalar field inside and outside the star in the $f(R)$ model (\ref{staro}) with $n=1$. 
A key parameter is the ratio between the energy density at infinity (i.e. the cosmological energy density) and the energy density at the center of the star, which can be 
parametrized by
$
v_0=M_P^2\R0/\trho_c.
$
Realistic values of this parameter are extremely small and are numerically challenging to explore, because the scalar field value at the center is proportional to $v_0^3$, as we will see below. The smallest value we considered is $v_0=10^{-4}$, which is much higher than the realistic one. However we believe that the situation is qualitatively similar for smaller values.  
Instead of the shooting method used in \cite{KM1}, we have resorted to  a relaxation method. 
\begin{figure}[ht]
\centering
\includegraphics[width=0.48\textwidth, clip=true]{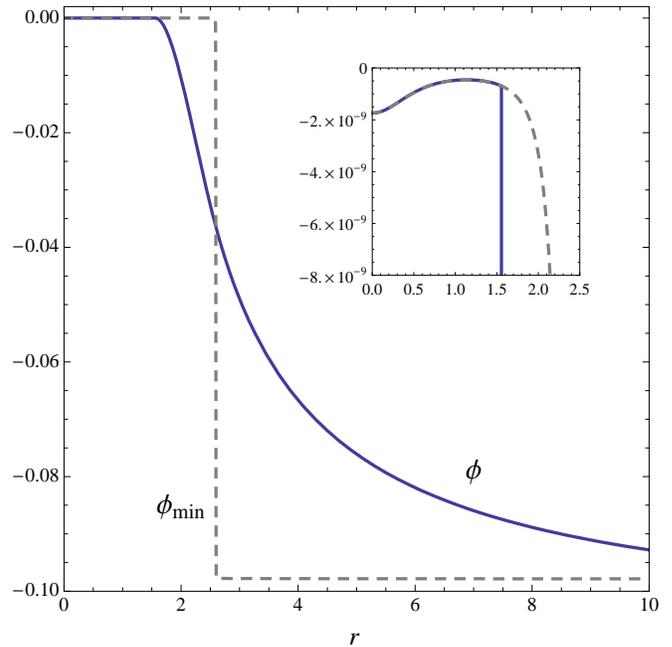}
\caption{Profile of  the scalar  field $\phi$ (in Planck units), shown by solid (blue) line, 
as a function of the radius (in units of $M_P\tilde{\rho}_c^{-1/2}$), 
for the model (\ref{staro}) with $n=1$, $\x1=3.6$ and $v_0=10^{-4}$. 
The value $\phi_{\rm min}$ for the minimum of the effective potential is plotted by dashed (gray) line.  }
\label{field}
\end{figure}

As is clear from Fig.~\ref{field}, the scalar field  tends to interpolate between an extremely high density regime, inside the star, and a very low density regime, outside the star. In the very high density regime, the scalar field is, numerically, very close to the singularity. This behaviour is quite analogous to that of the ``chameleon'' mechanism \cite{KW}, as we will discuss in more details in \cite{BL2}.

In order to understand intuitively our results, it is useful to introduce  the effective potential
\beq
\label{Veff}
V_{\rm eff}=V+\frac14 \Omega^4 (\trho-3\tP), 
\eeq
which is obtained by integrating, with respect to $\phi$,  the terms between brackets on the right hand side of the Klein-Gordon equation (\ref{KG}) 
(and we have set the constant of integration to zero).
The local minimum of  the effective potential, which we will denote $\phim(r)$, is characterized, if it exists, by
\beq
\frac{d V_{\rm eff}}{d\phi}=\frac{dV}{d\phi}+\frac{Q}{\mP}e^{4Q\phi/\mP}(\trho-3\tP)=0.\nonumber
\eeq
At high curvature ($x\gg 1$), $dV/d\phi\approx  \mP \R0\,  x/\sqrt{6}$, according to (\ref{approx}), and  the minimum corresponds to
\beq
\label{min}
x_{\rm min}\simeq \frac{\trho-3\tP}{M_P^2 \R0},\nonumber
\eeq
which is inversely proportional to $v_0$ (the scalar field value $\phi_{\rm min}$ is thus proportional to $v_0^3$).
This minimum exists only if the matter term $\trho-3\tP$ is positive. This implies that, for
a very compact constant density star, the effective potential in the central layers of the star does not have any mimimum. 

One can also define a corresponding effective squared mass as
\beq
\label{meff2}
m_{\rm eff}^2\equiv \frac{d^2V}{d\phi^2}+4\frac{Q^2}{\mP^2}e^{4Q\phi/\mP}(\trho-3\tP),
\eeq
which can become negative, as discussed earlier,  if the second term dominates and  $\trho-3\tP<0$. 
In the regime $x\gg 1$, if the minimum (\ref{min}) exists, one finds by evaluating (\ref{meff2}) at the minimum, 
using (\ref{approx}), 
\beq
m_{\rm eff}^2\approx \frac{\R0}{6\lambda n (2n+1)}x_{\rm min}^{2n+2}.\nonumber
\eeq

In view of our results, which seem contradictory with the conclusions of \cite{Frolov} and \cite{KM1}, it is worth reexamining  in detail their arguments against the existence of very relativistic stars. 
Frolov's argument is based on the non-relativistic limit of the Klein-Gordon equation which can be written in the 
form
\beq
\Delta \psi=-\frac{8\pi G}{3}\trho-2\frac{Q}{M_P}\frac{dV}{d\phi},\nonumber
\eeq
for $\psi=f_{\tR}-1\approx -2Q\phi/M_P$.
If one neglects the second term on the right hand side,  $\psi$ then satisfies an equation analogous to the Poisson 
equation $\Delta\Phi=4\pi G\rho$ and is thus related to the gravitational potential $\Phi$ by $\psi=\psi_\infty- 2\Phi/3$. 
Since $\psi_\infty$ is small, this suggests that the curvature singularity can be easily reached with a sufficiently strong gravitational potential. 
However, inside the star, the energy density is much higher than at infinity and this corresponds precisely to the high curvature regime, where $dV/d\phi$ becomes very high and cannot be neglected. 
\begin{figure}[h]
\centering
\includegraphics[width=0.4\textwidth, clip=true]{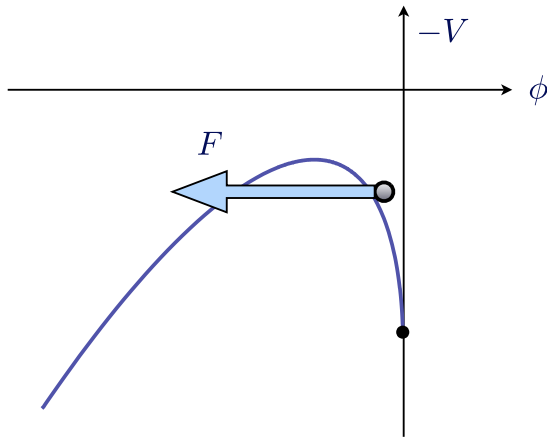}
\caption{Effective potential $U(\phi)=-V(\phi)$ in which the scalar field $\phi$ ``moves'' under the action of the force ${\cal F}$,
in the picture of the classical mechanic analogy. }
\label{analogy}
\end{figure}

One can also understand the scalar field profile, as in \cite{KM1}, in terms of a kinematic analogy  with the motion of a particle in a potential $U(\phi)$ which is submitted to a force ${\cal F}=-(8\pi G/3)(\trho-3 \tP)$, where the radius $r$ is considered to be a "fictive" time parameter (see Fig.~\ref{analogy}). The radial profile of the scalar field can then interpreted as the "motion" of the particle from $r=0$ to $r=\infty$. At $r=0$, the particle is at rest since $\phi'=0$, but is submitted to a leftward 
force that makes it roll uphill. The initial value for $\phi$ must be such that the scalar field reaches the top of the potential with vanishing velocity. 
We find a stronger gravitational potential in more compact stars with high energy density.  The amplitude of ${\cal F}$ then tends to become stronger, but there does not seem to be  any critical value beyond which the particle would necessarily fall into the singularity or overshoot the top of the hill. As the density of the star increases,
 the minimum $\phim$ of the effective potential  becomes closer to the singularity $\phi=0$, according to (\ref{min}). One thus expects   that the initial value of $\phi$ will also be closer and closer to the singularity 
$\phi=0$. This is confirmed by our numerical analysis, even if it becomes more and more difficult numerically to distinguish the extremely small value of $\phi$ from zero.

Moreover, as the numerical solution plotted in Fig.~\ref{field} shows explicitly, the ``motion" of the particle can be subtle. In this case, it first starts to approach the singularity, it then stops and moves and the opposite direction towards the top of the potential. This complicated ``motion" is a direct consequence of the non-monotonous dependence of $\trho-3\tP$, which reaches its maximum value at some non-zero radius of the star, as can be seen in Fig.~\ref{star}.

%\section{Conclusion}
To conclude, our work shows that it is possible to construct highly
relativistic stars in  $f(R)$ theories, despite recent indications to the
contrary.  Numerically, the task can be challenging as the scalar field
value is extremely close to the singularity in the center of the star. We
have also studied relativistic stars in chameleon-like models and found a
very similar behavior \cite{BL2}. In both cases, a crucial requirement for obtaining a static configuration is that the equation of state satisfies the condition $\trho-3\tP>0$ in most  of the star. Otherwise, the effective squared mass of the scalar field becomes negative and the associated instabilities prevent the construction of a static configuration. 
This is problematic for constant energy density stars, which have been
used in several recent works because of their analytical simplicity, since
$\trho-3\tP$ becomes negative in the central part of very massive stars of
this type.
This  might explain why, beyond some critical value for the gravitational field, no configuration was found in \cite{Tsujikawa:2009yf} for chamelon-like models. 
Realistic neutron stars do not, however, suffer from this problem.
Moreover, once the scalar field configuration is stable, the (hydrodynamical) stability of the stars should be very similar to that of their general relativistic counterparts because the backreaction of the scalar field on the star is extremely small.
An open question, which is far beyond the scope of this work, is whether the static configurations described here can be reached dynamically during the collapse of a very massive star into a neutron star. It would also be interesting 
to study whether and how black holes would form during the collapse of matter, i.e. such has been done for instance in \cite{Novak:1997hw} in the context of scalar-tensor theories without potential.

\vskip 1cm

{\bf Acknowledgements}
We would like to thank N.~Deruelle, G.~Esposito-Farese, J.~Novak, I.~Sawicki, A.~Starobinsky and S.~Tsujikawa for very instructive discussions. 
The work of E.B. was supported by the EU FP6 Marie Curie Research and Training Network ÒUniverseNetÓ (MRTN-CT-2006-035863).

\end{document}